\pdfoutput=1
\documentclass{article}
\usepackage{spconf,amsmath,amssymb,graphicx,booktabs,tabularx}
\usepackage{multirow,multicol}
\usepackage[table]{xcolor}
\usepackage{capt-of,balance}
\usepackage[hidelinks]{hyperref}
\DeclareMathSizes{9}{9}{7}{5}
\definecolor{Gray}{gray}{0.92}
\newcommand{\affilmark}[1]{\textsuperscript{\fontsize{9}{10}\selectfont#1}}
\newcommand{\modelname}{TF-Refiner}

\title{CONFIGURABLE-BANDWIDTH TIME--FREQUENCY MODELING FOR EFFICIENT FULL-BAND SPEECH ENHANCEMENT ACROSS SAMPLING RATES}
\hypersetup{
  pdftitle={CONFIGURABLE-BANDWIDTH TIME-FREQUENCY MODELING FOR EFFICIENT FULL-BAND SPEECH ENHANCEMENT ACROSS SAMPLING RATES},
  pdfauthor={Ui-Hyeop Shin, Wooseok Kim, Hyung-Min Park},
  pdfsubject={Full-band speech enhancement},
  pdfkeywords={speech enhancement, full-band audio, sampling-frequency independence, configurable computation}
}

\name{Ui-Hyeop Shin\affilmark{1,*}, Wooseok Kim\affilmark{2,*}, and
Hyung-Min Park\affilmark{1,2}
\thanks{\textsuperscript{*}Equal contribution. The order is alphabetical.}
\thanks{This work was partly supported by Institute of Information \& communications Technology Planning \& Evaluation (IITP) grant funded by the Korea government (MSIT) (RS-2022-II220989, Development of Artificial Intelligence Technology for Multi-speaker Dialog Modeling) and National Research Foundation of Korea (NRF) grant funded by the Korea government (MSIT) (RS-2026-25470024, Research on Spatial Audio Reasoning with Large Language Models across Diverse Microphone Arrays).}
}
\address{\affilmark{1}Department of Electronic Engineering, \affilmark{2}Department of Artificial Intelligence\\
Sogang University, Seoul, Republic of Korea\\
{\normalsize\{dmlguq123, wooseokkim, hpark\}@sogang.ac.kr}}

\begin{document}
\ninept
\normallineskiplimit=-3pt \lineskiplimit=-3pt
\raggedbottom
\setlength{\textfloatsep}{10pt plus 1pt minus 1pt}
\setlength{\dbltextfloatsep}{12pt plus 1pt minus 1pt}
\setlength{\abovedisplayskip}{6pt plus 1pt minus 1pt}
\setlength{\belowdisplayskip}{6pt plus 1pt minus 1pt}
\setlength{\abovedisplayshortskip}{3pt plus 1pt minus 1pt}
\setlength{\belowdisplayshortskip}{6pt plus 1pt minus 1pt}
\maketitle

\begin{abstract}

Speech enhancement systems are often developed for a fixed sampling rate, while time--frequency models become more expensive as the number of frequency bins increases. We propose \emph{TF-Refiner}, a sampling-frequency-independent model that decouples the deep analysis bandwidth from the full-band input and output. A deep encoder processes the band below a configurable cutoff, while a shallow decoder combines the encoded features with input-dependent high-band queries and predicts local complex filters applied to the original noisy STFT. A single parameter set trained at 16 and 48 kHz is evaluated at various sampling rates. On VoiceBank+DEMAND, the \emph{universal} model outperforms the rate-specific counterparts in PESQ, STOI, and log-spectral distance across the evaluated rates, including rates unseen in training. Random-cutoff training enables inference-time selection of cost--quality operating points without retraining or changing the output bandwidth. These results support configurable analysis bandwidth as a practical design choice for multi-rate full-band enhancement.

\end{abstract}


\vspace{-1mm}

\section{Introduction}
\label{sec:introduction}
\vspace{-1mm}

Speech enhancement (SE) is deployed at more than one sampling rate:
telephony and on-device pipelines operate at 16~kHz, whereas conferencing
and media applications increasingly require full-band 48-kHz
audio~\cite{dubey2022dns}, whose band above 8~kHz carries cues of
naturalness and intelligibility~\cite{stelmachowicz2001,monson2014,moore2016}.
In practice, a separate model is typically trained for each rate:
lightweight streaming models~\cite{gtcrn,rong2026ulunas,ahn2025fastenhancer} for
16~kHz and full-band systems~\cite{valin2018rnnoise,valin2020percepnet,schroter2022deepfilternet,schroter2022deepfilternet2,yu2022dmfnet}
for 48~kHz.
Sampling-frequency-independent (SFI) processing removes this duplication
in principle by tying the network to physical frequency rather than to a
fixed bin index~\cite{paulus2022sfi,saito2022sfi,zhang2023uses}, and
time--frequency (TF) dual-path
models~\cite{dang2022dptfsnet,wang2023tfgridnet,saijo2024tflocoformer},
which treat the frequency axis as a sequence, support this representation
naturally. However, existing SFI enhancement systems have relatively high computational demands~\cite{zhang2023uses,zhang2024uses2,fu2026onemodel}, and
the cost of a TF dual-path model grows almost
proportionally with the number of bins, so that the same architecture
becomes substantially more expensive at 48 than at 16~kHz~\cite{shin2026tfre}.

\begin{figure}[t]
  \centering
  \includegraphics[width=0.9\columnwidth]{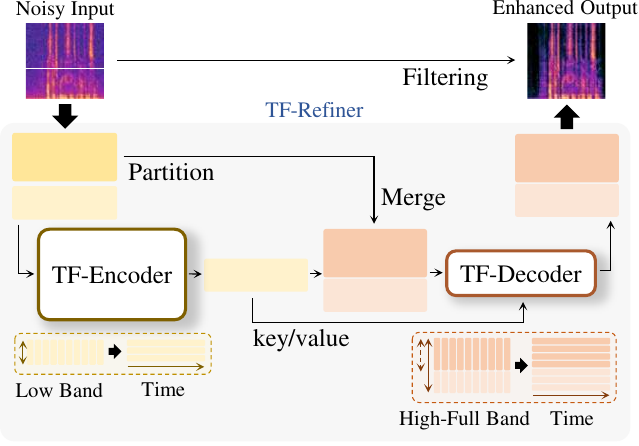}
  \vspace{-3mm}
  \caption{Overall flow of \modelname{}: deep analysis of the band below
  $f_c$, shallow full-band decoding with cross-attention to the encoder
  output, and filtering of the original noisy STFT.}
  \vspace{-1mm}
  \label{fig:concept}
\end{figure}

Practical full-band systems avoid this growth with hand-designed frequency
structures, such as perceptual-band envelopes~\cite{valin2018rnnoise,valin2020percepnet},
equivalent-rectangular-bandwidth (ERB) gains refined by complex filtering
only in the low band~\cite{schroter2022deepfilternet2}, or fixed sub-band
networks~\cite{yu2022dmfnet,hao2021fullsubnet}. In these designs, however,
the allocation of analysis effort across frequency is fixed at design time
and tied to one sampling rate.

These observations motivate a \emph{configurable analysis bandwidth}: a
single SFI model whose deep analysis is confined to a band defined in
physical frequency, while its full-band input and output follow the
sampling rate.
To this end, we propose \emph{\modelname}, an asymmetric encoder--decoder
for full-band SE, as shown in Figure~\ref{fig:concept}. It builds on the
query-based asymmetric structure of TF-Restormer~\cite{shin2026tfre}, which
analyzes only the observed band and synthesizes the unobserved band from
learnable extension queries, and moves the band partition from the edge of
the observed spectrum to a configurable cutoff $f_c$ inside it. A deep
TF-Encoder analyzes the band below $f_c$, whereas the complementary
observed band enters a shallow TF-Decoder as input-dependent queries that
attend to the encoder output~\cite{he2022mae,gupta2023siamese}, so that the
encoder cost is set by $f_c$ rather than by the sampling rate. Rather than
mapping the spectrum directly, the decoder predicts local complex filters
applied to the original noisy STFT~\cite{mack2020deepfiltering,shin2026ifcorrnet}, preserving
the observed resolution. Unlike the frequency projection of TF-Restormer,
which fixes a maximum bin count, no parameter of \modelname{} depends on
the number of bins, enabling one parameter set to serve any sampling rate.

Experiments show that a single \emph{universal} model trained at 16 and
48~kHz matches or exceeds rate-specific training in perceptual quality and
generalizes to unseen rates, while random-cutoff training lets $f_c$ set the
cost--quality trade-off at inference without retraining.

\section{TF-Refiner}
\label{sec:method}

\begin{figure}[t]
  \centering
  \includegraphics[width=0.88\columnwidth]{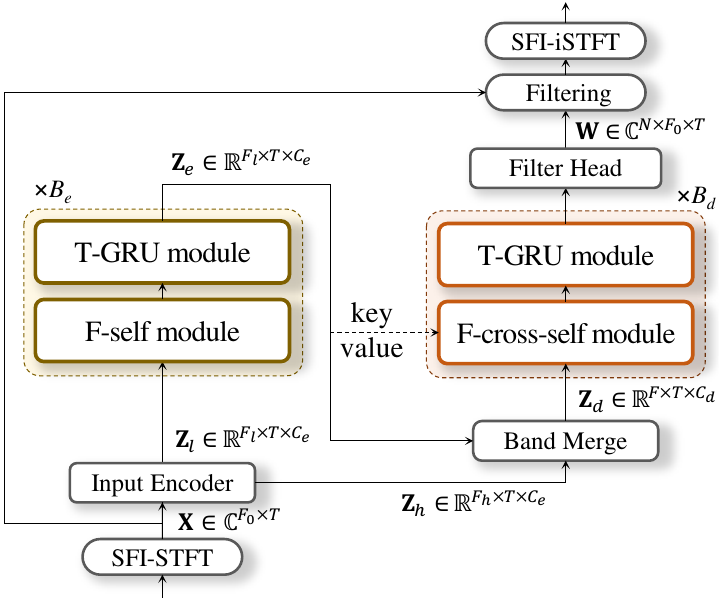}
  \vspace{-3mm}
  \caption{Architecture of \modelname{}. The $F_l$ positions below $f_c$ form $\mathbf{Z}_l$ for the
  deep TF-Encoder and the remaining $F_h$ positions form $\mathbf{Z}_h$. Band Merge concatenates $\mathbf{Z}_h$ with the encoder
  output $\mathbf{Z}_e$ for the shallow TF-Decoder, which also attends to $\mathbf{Z}_e$ as key/value.}
  \label{fig:overview}
  \vspace{-1mm}
\end{figure}

As shown in Figure~\ref{fig:overview}, \modelname{} realizes the
configurable bandwidth through an SFI representation
that keeps $f_c$ at the same physical frequency for every rate,
a deep TF-Encoder for the analysis band only, a shallow TF-Decoder
that refines the complementary observed band through cross-attention, and a Filter Head that predicts local
complex filters for the original noisy STFT. $F_l$ is fixed by $f_c$, so for a given cutoff
the sampling rate enters only through $F_h=F-F_l$.

\subsection{SFI representation and band partition}
\label{sec:sfi}

\begin{figure}[t]
  \centering
  \begin{minipage}[b]{0.3\columnwidth}\centering
    \includegraphics[width=\linewidth]{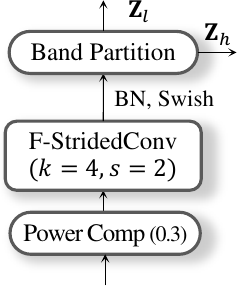}\\[2pt]{\small (a) Input Encoder}
  \end{minipage}\hfill
  \begin{minipage}[b]{0.32\columnwidth}\centering
    \includegraphics[width=\linewidth]{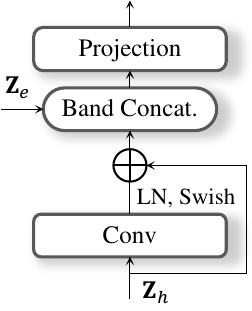}\\[2pt]{\small (b) Band Merge}
  \end{minipage}\hfill
  \begin{minipage}[b]{0.26\columnwidth}\centering
    \includegraphics[width=\linewidth]{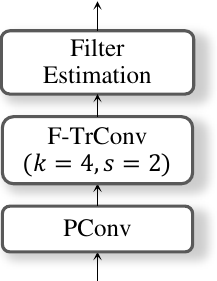}\\[2pt]{\small (c) Filter Head}
  \end{minipage}
  \vspace{-2mm}
  \caption{Components outside the TF modules.}
  \label{fig:parts}
\end{figure}

The SFI-STFT~\cite{paulus2022sfi,shin2026tfre} keeps a 40-ms square-root
Hann window and a 20-ms hop at every sampling rate $f_s$, so the bin
spacing is 25~Hz and the one-sided spectrum
$\mathbf{X}$ has $F_0=0.02f_s+1$ bins. The Input Encoder (Figure~\ref{fig:parts}(a))
compresses the magnitude as \mbox{$\mathbf{X}_c \hspace{-1mm}=\hspace{-1mm} |\mathbf{X}|^c \cdot e^{j\angle{\mathbf{X}}}$}, where $c=0.3$, and embeds the spectrum with a
frequency stride of 2, giving $\mathbf{Z}$
with $F=(F_0+1)/2$.
Band Partition splits the grid at the position of $f_c$ into $\mathbf{Z}_l$ and $\mathbf{Z}_h$ by $F_l=0.02f_c+1$.
We draw $f_c$ at random during training, so that one parameter set serves any
cutoff at inference (Section~\ref{sec:knob}).

\begin{figure}[t]
  \centering
  \begin{minipage}[b]{0.282\columnwidth}\centering
    \includegraphics[height=2.17in]{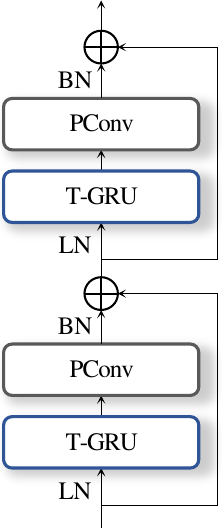}\\[1pt]{\small (a) T-GRU module}
  \end{minipage}
  \begin{minipage}[b]{0.282\columnwidth}\centering
    \includegraphics[height=2.17in]{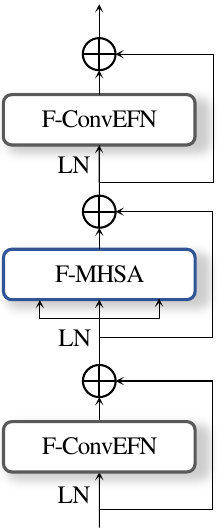}\\[1pt]{\small (b) F-self module}
  \end{minipage}
  \begin{minipage}[b]{0.412\columnwidth}\centering
    \includegraphics[height=2.17in]{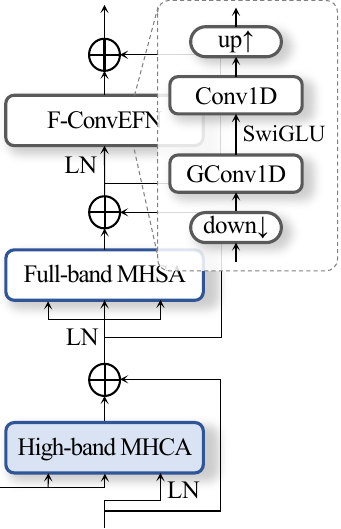}\\[1pt]{\small (c) F-cross-self module}
  \end{minipage}
  \vspace{-5mm}
  \caption{Unit modules of \modelname{} with the ConvEFN structure.}
  \label{fig:modules}
\end{figure}

\subsection{Asymmetric encoder--decoder}
\label{sec:asym}

The TF-Encoder applies $B_e$ blocks~\cite{dang2022dptfsnet,saijo2024tflocoformer}, each an F-self module based on multi-head self-attention (MHSA) with rotary position encoding~\cite{su2024roformer} and $A_e$ heads followed by
a T-GRU module (Figure~\ref{fig:modules}), to $\mathbf{Z}_l$ and yields $\mathbf{Z}_e$. Band Merge prepares the complementary band as
input-dependent queries through a residual convolution and normalization
(Figure~\ref{fig:parts}(b)), concatenates them with $\mathbf{Z}_e$ along
frequency, and projects the channels from $C_e$ to $C_d$ to form $\mathbf{Z}_d$.
The TF-Decoder applies $B_d$ blocks, each an F-cross-self module followed
by a T-GRU module. In the cross-self module only the $F_h$
complementary positions query $\mathbf{Z}_e$ through multi-head
cross-attention (high-band MHCA) with $A_d$ heads; the low-band positions bypass it. A
full-band MHSA with $A_d$ heads over all $F$ positions then
relates the two bands.
Both modules follow TF-Restormer~\cite{shin2026tfre} and use convolutional
efficient feed-forward networks (ConvEFN) on a length-reduced
sequence~\cite{shin2024sepreformer}: MHSA between two ConvEFNs in the F-self
module, and high-band MHCA, full-band MHSA, and one ConvEFN in sequence in the
F-cross-self module, with hidden widths $H_e$ and $H_d$ and two-group
convolutions.
The T-GRU module stacks two unidirectional GRU residual units.

\subsection{Deep filtering}
\label{sec:filter}

Let $X_{tf}\in\mathbb{C}$ denote the noisy STFT coefficient at frame $t$ and bin $f$, and let $\mathbf{x}_{tf}\in\mathbb{C}^{N}$ collect its $N$ neighboring coefficients in time and frequency. 
The Filter Head (Figure~\ref{fig:parts}(c)) mixes the decoder channels with a pointwise convolution, restores the STFT frequency resolution with a transposed convolution along frequency, and estimates
the complex filter tensor $\mathbf{W}$ with a convolutional layer. Following the three-component estimation structure of
SR-CorrNet~\cite{shin2026srcorrnet}, three real vectors per bin
$\mathbf{m}_{tf},\mathbf{a}_{tf},\mathbf{b}_{tf}\in\mathbb{R}^{N}$ give
$\mathbf{w}_{tf}=\operatorname{softplus}(\mathbf{m}_{tf})\odot[\tanh(\mathbf{a}_{tf})+j\tanh(\mathbf{b}_{tf})]$.
As in deep
filtering~\cite{mack2020deepfiltering,shin2026ifcorrnet}, the
enhanced STFT is $\widehat S_{tf}=\mathbf{w}_{tf}^{T}\mathbf{x}_{tf}$, and
the SFI-iSTFT reconstructs the waveform. The compressed features determine
the filter, but the samples being combined are the uncompressed
observations.

\begin{table}[t]
  \centering
  \vspace{-2mm}
  \caption{The universal model and the two rate-specific models on VBD at the default cutoff $f_c=5$~kHz (*: rates absent from training).}
  \label{tab:rates}
  \vspace{3pt}
  \setlength{\tabcolsep}{2pt}
  \renewcommand{\arraystretch}{0.98}
  \scalebox{0.88}{%
  \begin{tabular}{@{}cclccccc@{}}
    \toprule
    \textbf{Rate (kHz)} & \textbf{MACs} & \textbf{Model} & \textbf{PESQ$^{\uparrow}$} & \textbf{STOI$^{\uparrow}$} & \textbf{SI-SDR$^{\uparrow}$} & \textbf{LSD$^{\downarrow}$} & \textbf{LSD$_{>8\mathrm{k}}^{\downarrow}$} \\
    \midrule
    \multirow{4}{*}{16} & -- & Noisy & 1.97 & 0.921 & 8.47 & 1.193 & -- \\
     & \multirow{3}{*}{2.17G} & Universal & \textbf{3.46} & \textbf{0.952} & \textbf{19.69} & \textbf{0.619} & -- \\
     &  & 16k only & 3.35 & 0.950 & 19.20 & 0.624 & -- \\
     &  & 48k only$^*$ & 3.03 & 0.928 & 18.08 & 0.804 & -- \\
    \midrule
    \multirow{4}{*}{24*} & -- & Noisy & 1.97 & 0.921 & 8.44 & 1.297 & 1.349 \\
     & \multirow{3}{*}{2.63G} & Universal & \textbf{3.44} & \textbf{0.952} & \textbf{19.62} & \textbf{0.651} & 0.560 \\
     &  & 16k only & 3.34 & 0.949 & 19.13 & 0.660 & \textbf{0.549} \\
     &  & 48k only & 3.18 & 0.939 & 18.58 & 0.758 & 0.702 \\
    \midrule
    \multirow{4}{*}{32*} & -- & Noisy & 1.97 & 0.921 & 8.43 & 1.288 & 1.269 \\
     & \multirow{3}{*}{3.14G} & Universal & \textbf{3.47} & \textbf{0.953} & \textbf{19.47} & \textbf{0.660} & \textbf{0.595} \\
     &  & 16k only & 3.31 & 0.949 & 19.07 & 0.691 & 0.628 \\
     &  & 48k only & 3.29 & 0.946 & 18.60 & 0.725 & 0.691 \\
    \midrule
    \multirow{4}{*}{44.1*} & -- & Noisy & 1.97 & 0.921 & 8.42 & 1.314 & 1.290 \\
     & \multirow{3}{*}{4.03G} & Universal & \textbf{3.47} & \textbf{0.952} & 18.27 & \textbf{0.721} & \textbf{0.705} \\
     &  & 16k only & 3.23 & 0.949 & \textbf{18.96} & 0.748 & 0.722 \\
     &  & 48k only & 3.34 & 0.949 & 17.61 & 0.751 & 0.752 \\
    \midrule
    \multirow{4}{*}{48} & -- & Noisy & 1.97 & 0.921 & 8.41 & 1.314 & 1.285 \\
     & \multirow{3}{*}{4.35G} & Universal & \textbf{3.47} & \textbf{0.951} & 18.27 & \textbf{0.729} & \textbf{0.720} \\
     &  & 16k only$^*$ & 3.22 & 0.949 & \textbf{18.99} & 0.785 & 0.777 \\
     &  & 48k only & 3.35 & 0.949 & 17.74 & 0.735 & 0.730 \\
    \bottomrule
  \end{tabular}}
\end{table}

\begin{figure*}[t]
  \centering
  \includegraphics[width=0.999\textwidth]{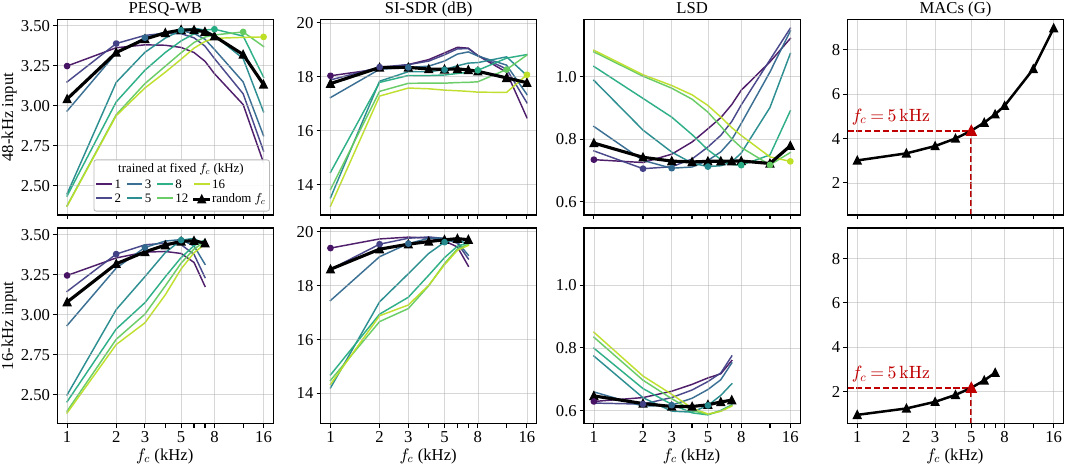}
   \vspace{-5mm}
 \caption{Changing the analysis cutoff $f_c$ at inference, for 48-kHz
  (top) and 16-kHz input (bottom). Colored curves: models trained with
  $f_c$ fixed, one color per training cutoff, the circle marking that
  cutoff (capped at 7~kHz for 16-kHz input, hence no circle for the 8-, 12-,
  and 16-kHz models there). Black: the universal model trained with a randomly
  drawn cutoff. All curves come from one checkpoint per model evaluated at each
  $f_c$; the cost depends only on the input rate and $f_c$, and red marks the
  default $f_c=5$~kHz of Table~\ref{tab:rates}.}
  \vspace{-1mm}
  \label{fig:cutoff}
\end{figure*}

\section{Experimental setup}
\label{sec:setup}

\subsection{Datasets and evaluation metrics}

We evaluate on the 824-utterance VoiceBank+DEMAND (VBD) test set at 16 and
48~kHz~\cite{valentini2017vbd}; its fixed mixtures use SNRs of
$\{2.5,7.5,12.5,17.5\}$~dB. The 16-kHz data are the official
28-speaker VBD training pairs at $\{0,5,10,15\}$~dB, divided into
10,413 training and 1,159 validation utterances. The 48-kHz training data mix clean VCTK
speech~\cite{veaux2017vctk} from 98 training and seven validation
speakers, excluding both test speakers, with DNS full-band
noise~\cite{dubey2022dns} at an integer SNR drawn uniformly from
$\{0,\ldots,19\}$~dB. VCTK utterances overlapping the VBD validation set are removed
from the 48-kHz pool.

We report wideband PESQ~\cite{rix2001pesq}, STOI~\cite{taal2011stoi},
SI-SDR~\cite{leroux2019sisdr}, and log-spectral distance (LSD) over the full
input band and above 8~kHz (LSD$_{>8\mathrm{k}}$, up to the input Nyquist
frequency and hence undefined at 16~kHz).
Inference runs at the input rate; SI-SDR and LSD are computed at that rate against the
original clean signal, whereas PESQ and STOI are computed after resampling to 16~kHz.
For the multi-rate evaluation the same 48-kHz test pairs are resampled to 16, 24,
32, and 44.1~kHz. We measure multiply--accumulate operations (MACs) for one second of input
using ptflops.

\subsection{Training and model configuration}

We train the universal model by alternating 16-kHz VBD and 48-kHz VCTK+DNS batches. 
We use 20 epochs of 20,000 minibatches each (400,000 in total), batch size 2,
2-s crops, and AdamP~\cite{heo2021adamp} (weight decay 1e-2,
$\beta=(0.95,0.999)$). The learning rate is warmed up for 1,000 steps
to 2e-3 and then cosine-decayed to 1e-6. Training draws cutoffs uniformly
from $\{2,\dots,8,12,16\}$~kHz for 48-kHz batches and
$\{2,\dots,7\}$~kHz for 16-kHz batches. One checkpoint per model is used for all reported input rates and inference cutoffs.
We use the objective of FastEnhancer~\cite{ahn2025fastenhancer},
$\mathcal L=0.3\mathcal L_{\rm mag}+0.2\mathcal L_{\rm cplx}+0.3\mathcal
L_{\rm cons}+0.2\mathcal L_{\rm wav}+0.001\mathcal L_{\rm pesq}$,
with power-compressed spectral terms.
We use the \modelname{}-B configuration in Table~\ref{tab:channel-config}.
The model applies deep filtering over a $3\times3$ TF neighborhood ($N=9$: one
past and one future frame, $\pm1$ bin), with $3\times3$ convolutions in
Band Merge and filter prediction. Each convolution adds one future frame,
giving a 40-ms look-ahead and, with the 40-ms window, an 80-ms algorithmic latency.
The default cutoff is $f_c=5$~kHz.

\section{Results}
\label{sec:results}

\subsection{Universal enhancement across sampling rates}

Table~\ref{tab:rates} evaluates the universal model trained at 16 and 48~kHz. 
The two rate-specific counterparts are trained only on data at their respective rate.
The universal model maintains comparable perceptual quality at the unseen rates and improves every metric over the noisy input. It obtains the best PESQ, STOI, and LSD at every rate and the best SI-SDR up to 32~kHz, whereas the 16-kHz model keeps the highest SI-SDR at 44.1 and 48~kHz, presumably because its training is confined to the power-dominant low band. 
The rate-specific models also transfer to unseen input rates without adaptation, although their PESQ decreases gradually.
With the default cutoff of $f_c=5$~kHz, the computational cost grows by
about 2.0 times from 16 to 48~kHz, whereas the number of bins triples, because the encoder analysis band remains fixed while only the shallow decoder processes
the wider observed band.

Because the VBD speech is drawn from VCTK, the two training streams do
not provide disjoint clean-speech corpora; they differ mainly in sampling
bandwidth and corruption generation. Combining VBD pairs with
VCTK+DNS mixtures may therefore contribute to the universal
model's advantage, but the 48-kHz-only model already uses the larger VCTK+DNS pool yet remains below
the universal model at its native rate. We note that Table~\ref{tab:rates} does
not isolate the contributions of data diversity and joint
sampling-rate exposure.

\begin{table}[t]
  \centering
  \caption{Results on VBD at 16 and 48~kHz. Each \modelname{} variant uses
  the same parameter set for both rates.
  GTCRN and UL-UNAS report SI-SNR. DeepFilterNet1/2~\cite{schroter2022deepfilternet, schroter2022deepfilternet2} use DNS training data.  
  *: zero-shot results for unseen rates. $^\dagger$: no-look-ahead version.}
  \label{tab:main}
  \vspace{3pt}
  \setlength{\tabcolsep}{1.5pt}
  \renewcommand{\arraystretch}{1.05}
  \scalebox{0.88}{%
  \begin{tabular}{@{}lccccccc@{}}
    \toprule
    \multirow{2}{*}{\textbf{Method}} & \multirow{2}{*}{\textbf{Param}} & \multicolumn{3}{c}{\textbf{16 kHz}} & \multicolumn{3}{c}{\textbf{48 kHz}} \\[-1pt]
    \cmidrule(lr){3-5} \cmidrule(l){6-8}
     & & \textbf{MACs} & \textbf{PESQ} & \textbf{SI-SDR}
           & \textbf{MACs} & \textbf{PESQ} & \textbf{SI-SDR} \\
    \midrule
    Noisy                                             & --    & --    & 1.97  & 8.47  & --    & 1.97 & 8.41  \\
    \midrule
    GTCRN~\cite{gtcrn}                              & 0.02M & 0.04G & 2.87  & 18.83 & --    & --  & --    \\
    UL-UNAS~\cite{rong2026ulunas}                   & 0.17M & 0.03G & 3.09  & 18.48 & --    & --  & --    \\
    FastEnhancer-B~\cite{ahn2025fastenhancer}       & 0.09M & 0.26G & 3.13  & 19.0  & --    & --  & --    \\
    FastEnhancer-S~\cite{ahn2025fastenhancer}       & 0.20M & 0.66G & 3.19  & 19.2  & --    & --  & --    \\
    FastEnhancer-M~\cite{ahn2025fastenhancer}       & 0.49M & 2.90G & 3.24  & 19.4 & --    & --  & --    \\
    \rowcolor{Gray} \modelname{}-B (\textit{16k}) & 0.47M & 2.17G & 3.35 & 19.20 & 4.35G & 3.22*\hspace{-1.2mm} & \bf 18.99*\hspace{-1.2mm} \\
    \midrule
    PercepNet~\cite{valin2020percepnet}            & 8.00M & --    & --    & --    & 0.80G & 2.73 & --    \\
    DeepFilterNet~\cite{schroter2022deepfilternet}  & 1.78M & --    & --   & --    & 0.35G & 2.81 & 16.63 \\
    DMF-Net~\cite{yu2022dmfnet}                     & 7.84M & --    & --   & --    & --    & 2.97  & --    \\
    DeepFilterNet2~\cite{schroter2022deepfilternet2}& 2.31M & --    & --   & --    & 0.36G & 3.08  & --    \\
    \rowcolor{Gray} \modelname{}-B (\textit{48k}) & 0.47M & 2.17G & 3.03*\hspace{-1.2mm} & 18.08*\hspace{-1.2mm} & 4.35G & 3.35 & 17.74 \\
    \midrule
    \rowcolor{Gray} \modelname{}-B & 0.47M & 2.17G & \textbf{3.46} & \textbf{19.69} & 4.35G & \textbf{3.47} & 18.27 \\
    \rowcolor{Gray} \modelname{}-T$^\dagger$ & 0.09M & 0.54G & 3.20 & 19.07 & 1.27G & 3.10 & 17.07 \\
    \rowcolor{Gray} \modelname{}-S$^\dagger$ & 0.19M & 0.97G & 3.37 & 19.42 & 2.03G & 3.36 & 18.36 \\
    \rowcolor{Gray} \modelname{}-B$^\dagger$ & 0.45M & 2.05G & 3.30  & 19.42 & 3.85G & 3.36 & 18.41 \\
    \bottomrule
  \end{tabular}}
  \vspace{-1mm}
\end{table}

\subsection{Configurable computation}
\label{sec:knob}

Figure~\ref{fig:cutoff} compares the universal model with fixed-cutoff
controls while changing $f_c$ at inference for every model. The controls
use the same 16/48-kHz training with $f_c$ fixed at 1, 2, 3, 5, 8, 12,
or 16~kHz. 
The total MACs increase with $f_c$ because more positions enter the deep encoder.
Fixed-cutoff models degrade as the inference cutoff moves far from their training cutoff. Random-cutoff training keeps
the universal model close to the best fixed-cutoff model in PESQ and LSD over 2--8~kHz
(2--7~kHz for 16-kHz input),
while the fixed-cutoff models retain an advantage at the extremes.

Lowering $f_c$ from the default 5 to 3~kHz reduces the cost by 29\% at
16~kHz and 16\% at 48~kHz with a small PESQ decrease. At lower cutoffs,
the universal model also degrades more gradually than the model trained
at a fixed 5-kHz cutoff. At 48~kHz, the model trained and evaluated at $f_c=16$~kHz costs about twice as much as the universal model at the default cutoff without a higher PESQ.

\begin{table}[t]
  \centering
  \caption{Model configurations of the \modelname{} variants.}
  \vspace{2mm}
  \label{tab:channel-config}
  \setlength{\tabcolsep}{5pt}
  \renewcommand{\arraystretch}{0.95}
  \scalebox{0.9}{%
  \begin{tabular}{@{}lcccccccc@{}}
    \toprule
    \multirow{2}{*}{\textbf{Model}}&\multicolumn{4}{c}{\textbf{Encoder}}&\multicolumn{4}{c}{\textbf{Decoder}}\\
    \cmidrule(lr){2-5} \cmidrule(l){6-9}
     & \textbf{$B_e$} & \textbf{$C_e$} & \textbf{$H_e$} & \textbf{$A_e$}
    & \textbf{$B_d$} & \textbf{$C_d$} & \textbf{$H_d$} & \textbf{$A_d$} \\
    \midrule
    TF-Refiner-T & 4 & 24 & 42  & 4 & 2 & 12 & 24 & 2  \\
    TF-Refiner-S & 4 & 32 & 88  & 4 & 2 & 16 & 48 & 4  \\
    TF-Refiner-B & 4 & 48 & 144 & 4 & 2 & 24 & 72 & 4  \\
    \bottomrule
  \end{tabular}}
  \vspace{-3mm}
\end{table}

\subsection{Comparison with published systems}
\label{sec:benchmark}

Table~\ref{tab:main} compares published systems at their native rates with \modelname{} variants, each using one parameter set at both rates. 
For single-rate comparisons, we include \modelname{}-B (\textit{16k}) and \modelname{}-B (\textit{48k}), the rate-specific models of Table~\ref{tab:rates}. At the default $f_c=5$~kHz, \modelname{}-B (\textit{16k})
achieves a PESQ of 3.35 versus 3.24 for FastEnhancer-M at 16~kHz, while its
SI-SDR is 0.2~dB lower. \modelname{}-B (\textit{48k}) achieves a PESQ of 3.35
versus 3.08 for DeepFilterNet2~\cite{schroter2022deepfilternet2} at 48~kHz.
The universal \modelname{}-B further increases PESQ to 3.46 and 3.47 at
16 and 48~kHz, respectively.

The no-look-ahead variants ($^\dagger$) use frequency-only convolutions with kernel size 3 in Band Merge and filter prediction, with $N=3$ same-frame frequency taps, giving a 40-ms algorithmic latency.
\modelname{}-B$^\dagger$ also outperforms the published systems in PESQ at both rates.
For scaling, we evaluate the tiny and small no-look-ahead variants using the configurations in Table~\ref{tab:channel-config}. S$^\dagger$ maintains comparable quality to B$^\dagger$ at substantially lower cost, while T$^\dagger$ trades some quality for further savings.

\subsection{Ablation study}

\begin{table}[t]
  \centering
  \caption{Ablation of the rate-specific 48-kHz model.}
  \label{tab:ablation}
  \vspace{3pt}
  \setlength{\tabcolsep}{3pt}
  \renewcommand{\arraystretch}{1.05}
  \scalebox{0.9}{%
  \begin{tabular}{@{}lccccc@{}}
    \toprule
    \textbf{Model} & \textbf{PESQ$^{\uparrow}$} & \textbf{STOI$^{\uparrow}$} & \textbf{SI-SDR$^{\uparrow}$} & \textbf{LSD$^{\downarrow}$} & \textbf{LSD$_{>8\mathrm{k}}^{\downarrow}$} \\
    \midrule
    \rowcolor{Gray} \modelname{}-B (\textit{48k}) & \textbf{3.35} & 0.949 & \textbf{17.74} & \textbf{0.735} & \textbf{0.730} \\
    w/o high-band MHCA      & 3.31 & \textbf{0.950} & 17.40 & 0.782 & 0.794 \\
    w/o Band Merge conv.    & 3.34 & \textbf{0.950} & 17.39 & 0.767 & 0.773 \\
    w/ direct mapping       & 3.11 & 0.944 & 15.95 & 0.801 & 0.812 \\
    \bottomrule
  \end{tabular}}
\end{table}

Table~\ref{tab:ablation} ablates the modules that connect the analysis band to the full-band output.
Removing high-band MHCA, through which the high-band queries attend to the encoder output, or the Band Merge convolution that forms those queries changes PESQ and STOI, which assess only the band below 8~kHz, by at most 0.04 but costs about 0.3~dB of SI-SDR and raises LSD$_{>8\mathrm{k}}$ from 0.730 to 0.794 and 0.773, respectively. Both modules thus mainly serve the observed high band.
Replacing local filtering with direct complex mapping degrades every metric, lowering SI-SDR by 1.8~dB and raising LSD by 0.07, consistent with the filter preserving the fine structure of the observed STFT.

\section{Conclusion}
\label{sec:conclusion}

We presented \modelname{}, a sampling-frequency-independent full-band
speech enhancement model whose deep analysis is confined to a
configurable band while a shallow decoder refines the observed remainder
and filters the original noisy STFT. One universal model enhances 16- to 48-kHz inputs, including unseen rates, and the analysis
cutoff selects a cost--quality operating point at inference without changing
the architecture or retraining. Future work includes evaluation on unseen
corpora and streaming latency measurement.

\clearpage
\small
\let\origthebibliography\thebibliography
\def\thebibliography#1{\origthebibliography{#1}\setlength{\itemsep}{0pt}\setlength{\parskip}{0pt}}
\bibliographystyle{IEEEbib_compact}
\bibliography{references}

\end{document}